\documentclass[11pt, oneside]{article}   	
\usepackage{geometry}                		
\usepackage{graphicx}	
\newcommand\ignore[1]{}			
\usepackage{amssymb}
\usepackage{amsmath}
\usepackage{amssymb}
\usepackage{amsthm}
\usepackage{graphicx}
\usepackage[urlcolor=blue,colorlinks=true]{hyperref}
\hypersetup{colorlinks,linkcolor={blue},citecolor={blue},urlcolor={red}}
\usepackage{placeins}
\usepackage{tocloft}
\usepackage{slashed}
\usepackage{float}
\usepackage{fullpage}
\usepackage{braket}
\usepackage{url}

\usepackage{tabularx}

\usepackage{cite}
\usepackage{subfig}
\usepackage{upgreek}
\usepackage{outlines}
\usepackage{braket}
\usepackage{appendix}
\def\0{{(0)}}
\def\1{{(1)}}
\usepackage{xspace}

\def\bbP{\mathbb{P}}

\def\d{\partial}

\def\a{\alpha}
\def\b{\beta}

\usepackage{bm}

\def\g{\gamma}

\usepackage{caption}

\def\x{\chi}

\def\t{\tau}

\def\Om{\Omega}

\def\l{\lambda}

\def\onec{\left(\begin{array}{c}}
\def\cend{\end{array}\right)}
\def\cc{\left(\begin{array}{cc}}
\def\ccend{\end{array}\right)}
\def\ccc{\left(\begin{array}{ccc}}

\def\cccend{\end{array}\right)}
\def\cccc{\left(\begin{array}{cccc}}
\def\cccc{\left(\begin{array}{ccccc}}

\def\ccccend{\end{array}\right)}

\def\t{\tau}

\renewcommand\mod[1]{\text{ mod }#1}

\usepackage{tikz}
\usepackage{pgfplots}
\usetikzlibrary{matrix,arrows,decorations.pathmorphing}
\usetikzlibrary{patterns}
\usetikzlibrary{shapes.misc}
\usetikzlibrary{trees}
\usetikzlibrary{decorations.pathmorphing}
\usetikzlibrary{shapes.geometric}
\usetikzlibrary{positioning}
\usetikzlibrary{decorations.markings}
\usetikzlibrary{positioning,arrows}
\newcommand\numberthis{\addtocounter{equation}{1}\tag{\theequation}}
\usetikzlibrary{decorations.pathreplacing}
\usetikzlibrary{decorations.pathmorphing}
\usetikzlibrary{shapes}
\usetikzlibrary{arrows.meta}
\usetikzlibrary{plotmarks}
\usetikzlibrary{decorations.markings}
\tikzset{
particle/.style={thin,draw=black, postaction={decorate},
decoration={markings,mark=at position .5 with {\arrow[black, line width=0.5mm]{stealth}}}},
gluon/.style={decorate, draw=black, decoration={coil,amplitude=4pt, segment length=5pt}},
photon/.style={decorate, decoration={snake}},
singularity/.style={decorate, draw=black, decoration=zigzag}
}

\newcommand{\eq}[1]{\begin{align}#1\end{align}}
\newcommand{\seq}[1]{\begin{align*}#1\end{align*}}
\newcommand{\subeqs}[1]{\begin{subequations}\begin{align}#1\end{align}\end{subequations}}

\def\de{\delta}
\def\De{\Delta}

\usepackage[bottom]{footmisc}

\newcommand{\sltz}{\SL(2,\Z)}

\theoremstyle{definition}

\usepackage{amsthm}
\theoremstyle{theorem}

\def\bbQ{\mathbb{Q}}
\def\bbZ{\mathbb{Z}}
\def\calB{\mathcal{B}}
\def\bbC{\mathbb{C}}

\def\bbP{\mathbb{P}}
\def\xfb{\chi_{4.b}}

\def\sltz{\operatorname{SL}(2,\bbZ)}
\usepackage{authblk}
\numberwithin{equation}{section}
\def\bbH{\mathbb{H}}
\def\hflux{H_{\text{flux}}}

\def\e{\mathcal{E}}
\def\calE{\mathcal{E}}
\def\ep{\e_\phi}
\def\epp{\e'_\phi}
\def\x{\chi}

\title{Flux Modularity, F-Theory, and Rational Models}
\author{Shamit Kachru, Richard Nally, and Wenzhe Yang}
\affil{Stanford Institute for Theoretical Physics,\\ Stanford University, Stanford, CA, 94305}
\begin{document}
\maketitle

\begin{abstract}
In recent work, we conjectured that Calabi-Yau threefolds defined over $\mathbb{Q}$ and admitting a supersymmetric flux compactification are modular, and associated to (the Tate twists of) weight-two cuspidal Hecke eigenforms. In this work, we will address two natural follow-up questions, of both a physical and mathematical nature, that are surprisingly closely related. First, in passing from a complex manifold to a rational variety, as we must do to study modularity, we are implicitly choosing a ``rational model" for the threefold; how do different choices of rational model affect our results? Second, the same modular forms are associated to elliptic curves over $\mathbb{Q}$; are these elliptic curves found anywhere in the physical setup? By studying the F-theory uplift of the supersymmetric flux vacua found in the compactification of IIB string theory on (the mirror of) the Calabi-Yau hypersurface $X$ in $\mathbb{P}(1,1,2,2,2)$, we find a one-parameter family of elliptic curves whose associated eigenforms exactly match those associated to $X$. Actually, we find two such families, corresponding to two different choices of rational models for the same family of Calabi-Yaus.  
\end{abstract}
\clearpage
\tableofcontents
\section{Introduction}
One of the crowning achievements of twentieth century mathematics was the modularity theorem \cite{Wiles,Taylor2,Diamond,Conrad,Breuil}, which implies that every elliptic curve defined over $\bbQ$ can be associated to a special automorphic form called a weight-two eigenform. More recently, the modularity program has been extended to higher-dimensional Calabi-Yau (CY) varieties, including CY threefolds (CY3s) \cite{yui:review,meyer:book}. For example, it is known that rigid threefolds defined over $\bbQ$ are modular \cite{yui:rigid}. Recently, the modularity of threefolds has been connected to special string compactifications. In particular, it has been conjectured that supersymmetric flux compactifications over $\bbQ$ are associated to the same kind of modular forms as elliptic curves \cite{RAN:fluxModularity}, and that rank-two attractors defined over $\bbQ$ are associated to the same kind of modular form as rigid threefolds \cite{candelas:attractors}.

Although these relationships were in some sense anticipated in \cite{moore:a&along,moore:arithmeticLectures}, for technical reasons they are somewhat surprising. In physics, when we consider a string compactification we usually think of the threefold as being more or less uniquely specified by the choice of complex structure. However, to discuss modularity this is insufficient; we must also choose a rational model for the threefold. Indeed, the eigenforms associated to elliptic curves and threefolds by the various modularity theorems and conjectures are most properly thought of as being assigned to this choice of rational model, and the same point in complex structure moduli space can be related to many different eigenforms. Thus, a deep understanding of the relationship between string theory and modularity will require significant conceptual and technical advances beyond what is currently understood.

The main question that motivates this paper is why, physically, supersymmetric flux compactifications are modular. We will not answer this question, but we will provide a physical picture that at least makes the results of \cite{RAN:fluxModularity} plausible. Weight-two eigenforms, which are the automorphic objects that we associated to supersymmetric flux compactifications in \cite{RAN:fluxModularity}, are more traditionally associated to elliptic curves defined over $\bbQ$. With this in mind, it is natural to ask if there is a rational elliptic curve anywhere in the physical setup. 

On the other hand, it is well-known that flux compactifications of IIB string theory on a CY3 are dual to F-theory compactified on an elliptically-fibered CY fourfold \cite{sen:orientifold}. F-theory is an approach to IIB string theory that geometrizes certain non-perturbative aspects \cite{vafa:ftheory,morrison:ftheory1,morrison:ftheory2}; see e.g. \cite{weigand:review} for a review. In the F-theory uplift of the flux compactification, the generic elliptic fiber\footnote{Throughout this paper, we work in a limit where the axiodilaton $\tau$ is constant over the base of the elliptically fibered  fourfold.} provides a natural elliptic curve, which we will argue reproduces the modularity of the underlying flux compactification, as identified in \cite{RAN:fluxModularity}. 

Perhaps the most important challenge to this proposal is that a priori there is no reason for the generic elliptic fiber to be defined over $\bbQ$, even when the associated threefold is. Indeed, in F-theory the generic fiber is most naturally defined over the function field of the base, and thus does not have any obvious modular properties. Nevertheless, we will revisit the main example of \cite{RAN:fluxModularity} and find that, whenever the threefold is defined over $\bbQ$, so is the generic fiber of the F-theory uplift. Moreover, we will find that the framework of \cite{sen:orientifold} naturally produces a rational model for this elliptic curve, and that the modular form associated to this choice of rational model exactly matches the modular form that appears in the zeta-function of the threefold, subject to one choice of twist. In fact, we will study two different rational models for these flux compactifications, and correspondingly find two rational models for the associated family of elliptic curves, with matching in both cases\footnote{The discussion of different rational models for the Calabi-Yau hypersurface in $\bbP(1,1,2,2,2)$ and their modularity was motivated in part by a comment by one of the referees of \cite{RAN:fluxModularity}; we thank that referee for suggesting this.}.

For threefolds with $h^{2,1} = 1$, similar ideas appeared in Section 4 of \cite{candelas:attractors}. Our treatment extends that of \cite{candelas:attractors} in a few important respects. First, we describe the F-theory geometry explicitly. The authors of \cite{candelas:attractors} argued that in the case they study, F-theory gives an elliptic curve with the right $j$-invariant; here we exhibit the stronger result that, surprisingly, it also produces the desired rational model for an elliptic curve with this $j$-invariant. We are only able to see this by directly manipulating the defining equation of the F-theory fourfold. Additionally, whereas those authors considered an isolated flux compactification, we will consider a continuous family of vacua. Although this situation is less generic, it offers a rather richer laboratory to study flux modularity. To obtain a rational model from the F-theory construction, we must put in one rational model by hand; this allows us to pick out a coordinate transformation in the F-theory geometry that is necessary to correctly reproduce the modularity of the threefold. In a continuous family, this is not really an issue: we put in the rational model associated to one point in moduli space, and get out the rational models associated to an infinite number of other points. On the other hand, for isolated vacua this prescription has no predictive power: to correctly reproduce the rational model, we must insert the same rational model by hand. Put another way, while the authors of \cite{candelas:attractors} gave interesting evidence\textemdash in the form of the $j$-invariant\textemdash that the F-theory geometry should reproduce the weight-two eigenform associated to the flux compactification they described, we are able to show this explicitly in the setting we study, for all but one rational point along the family of flux compactifications.

The outline of this paper is as follows. First, in Section \ref{sec:twists}, we will briefly introduce the notion of rational models; these are unfamiliar to physicists, and are absolutely essential to modularity, so it is well worth taking the time to understand them. Next, in Section \ref{sec:sen} we will study Sen's construction of elliptically fibered fourfolds in F-theory, and obtain a universal formula for the generic fiber in terms of the $j$-invariant of the underlying supersymmetric flux compactification. Finally, in Section \ref{sec:main} we will apply the F-theory framework to the example of \cite{RAN:fluxModularity}, and derive two one-parameter families of elliptic curves that exactly reproduce the eigenforms associated to two families of rational models for the supersymmetric flux compactifications arising in the octic Calabi-Yau hypersurface in $\bbP(1,1,2,2,2)$. We will then conclude in Section \ref{sec:conclusion} with discussion and outlook. 

\section{Twists and Rational Models}
\label{sec:twists}
In string theory, we usually deal with CY3s in the category of complex manifolds, where (neglecting the K\"ahler structure) they are determined up to isomorphism solely by their complex structure. For instance, any two tori with the same $j$-invariant (or, equivalently, the same modular parameter $\tau$) are isomorphic as complex manifolds. However, to discuss modularity, we cannot work in the category of complex manifolds, but instead in the category of rational projective varieties, where the notion of isomorphism is much more refined. Put another way, to discuss the modularity of a variety, it is not enough to fix its complex structure; we must also pick a rational structure.

As an example, consider the two elliptic curves \subeqs{\calE: y^2 &= x^3 + x, \\\calE': y^2 &= x^3 - x.} These curves both have $j=1728$, so as complex tori they are isomorphic, but $\e$ has conductor 64 and $\e'$ has conductor 32.  The conductor determines the level of the modular forms associated to $\e$ and $\e'$, so as varieties defined over $\bbQ$ they are not isomorphic. This point of view is unfamiliar from the perspective of string theory, so we will spend this section briefly reviewing how different arithmetic objects can be isomorphic as complex manifolds, and the relationship between the modular forms associated to such objects. The material in this section is fairly standard, and much more detail can be found in e.g. \cite{silverman,silverman2,koblitz}.

\subsection{Rational Models}
\label{sec:rationalModels}
We begin with the elliptic curve \eq{\calE: y^2 = x^3 + fx + g,\label{eq:ECgeneral}} which has $j$-invariant \eq{j = 1728 \frac{4f^3}{4f^3+27g^2}.\label{eq:jGeneral}} The rescaling \eq{y\to\sqrt{d}y\label{eq:yRescale}} takes Eq. \ref{eq:ECgeneral} to \eq{\calE': y^2 = x^3 + d^2fx + d^3g,\label{eq:ECrescale}} which manifestly still has $j$-invariant given by Eq. \ref{eq:jGeneral}. If we are viewing the elliptic curves as complex tori, then Eq. \ref{eq:yRescale} is simply a change of variables, so Eqs. \ref{eq:ECgeneral} and \ref{eq:ECrescale} are isomorphic; this is simply the statement that all complex tori with fixed $j$ are isomorphic.  

Conversely, if we consider $\calE$ and $\calE'$ as being defined over $\bbQ$ (which of course requires $f,g,d\in\bbQ$) then the situation is somewhat more complicated. If $d$ is a perfect square, then $\sqrt{d}$ is also rational, and once again Eq. \ref{eq:yRescale} is a simple change of variables, so that Eqs. \ref{eq:ECgeneral} and \ref{eq:ECrescale} are isomorphic even over the rationals. 

On the other hand, if $d$ is not a perfect square, then the situation is radically different. In general, Eq. \ref{eq:yRescale} is not an isomorphism over $\bbQ$, but instead over the number field $\bbQ(\sqrt{d})$, and thus $\calE$ and $\calE'$ are not in general isomorphic over $\bbQ$. The elliptic curve $\calE'$ is called a quadratic twist of $\calE$ \cite{silverman,silverman2}, and in general $\calE$ and $\calE'$ are genuinely different objects in arithmetic geometry: they are different rational models for the same complex torus.

\subsection{Twists}
The weight-two eigenforms associated to two elliptic curves related by a quadratic twist are themselves simply related. Consider elliptic curves $\calE,\calE'$ whose associated eigenforms (called $f$  and $g$, respectively) have Fourier expansions given by \subeqs{f(\t) &= \sum_{n} a_nq^n, \\g(\t) &= \sum_{n}b_nq^n.} If $\e,\e'$ are related by a quadratic twist as above, then we will have \eq{a_p=\pm b_p\label{eq:ap=pmbp}} for all but finitely many primes $p$. 

The choice of sign is determined by a Dirichlet character. A Dirichlet character $\chi$ with modulus $D$ is a map $\chi:\bbZ\to\bbC$ satisfying the  equations \subeqs{\x(n) &= \x(n+D),\\\x(nm)&=\x(n)\x(m)} for all $n,m$; we say that $\x$ has order $k$ if $\x^k$ is a trivial character, i.e. if \eq{\x(n)^k=0\text{ or }1} for all $n$. Thus, if $\x$ is of order two we have $x(n)=0,\pm1$ for all $n$. More  detail about Dirichlet characters can be found in e.g. \cite{silverman,silverman2,koblitz}.

Quadratic fields $\bbQ(\sqrt{d})$ (where $d$ is square free) are in bijection with Dirichlet characters of order two and modulus $D$, where $D$ is determined by $d$ as \eq{D = \left\{\begin{array}{cc}4d & d\equiv 2,3\mod{4}\\d & d\equiv 1\mod{4}\end{array}\right..} Thus, the field over which a quadratic twist is defined is associated to a unique Dirichlet character $\x$; the more precise statement of Eq. \ref{eq:ap=pmbp} is that \eq{a_p = \x(p)b_p\label{eq:ap=xbp}} for all but finitely many primes $p$. Two modular forms related in this way are said to be twists of each other, and the modular forms of any two elliptic curves that are quadratic twists of each other will also be related by a twist.

Although we have given a precise definition of the map between modular forms related by a quadratic twist, we have not actually said how to evaluate the signs in Eq. \ref{eq:ap=xbp}. In this paper, we will focus on the Dirichlet character of modulus $D$ and order 2 \cite{montgomery_vaughan_2006}, which is given by \eq{\x(n) = \left(\frac{D}{n}\right),} where $\left(\frac{\cdot}{\cdot}\right)$ is a Kronecker symbol.

\subsection{A Suggestive Example}
\label{sec:xfb}
For an example of the above ideas that will be quite relevant later, consider the two elliptic curves \subeqs{y^2 &= x^3 + fx + g\\y^2 &= x^3 + fx - g,} where $f,g\in\bbQ$.  These two curves have the same $j$-invariant, but are arithmetically inequivalent, so they are different rational models for the same complex torus. To see that they are arithmetically inequivalent, note that the $g\to-g$ map is generated by the coordinate transformation \eq{y\to iy.\label{eq:ytoiy}} Thus, the two curves are isomorphic over $\bbQ(i)$, but not over $\bbQ$. 

What about the modular forms associated to these elliptic curves? By the discussion above, the two modular forms are related by a Dirichlet character for $\bbQ(i)$. The nontrivial character, which following LMFDB \cite{lmfdb} we will call $\xfb$, is defined  in terms of a Kronecker symbol as \eq{\xfb(n) = \left(\frac{-4}{n}\right) = \left\{\begin{array}{cc}0 & n\equiv 0\mod{4}\\ 1 & n\equiv 1\mod{4}\\0 & n\equiv 2\mod{4}\\-1 & n\equiv 3\mod{4}\end{array}\right. .\label{eq:xfb}} We have indicated its action on some small primes in Table \ref{tab:xfb}. Thus, any two elliptic curves related by Eq. \ref{eq:ytoiy} have the same point counts for all $p$ not congruent to 3 mod 4. 

\begin{table}[h]
\begin{centering}
\begin{tabular}{|c|c|c|c|c|c|c|}\hline
$p$ & 3 & 5 & 7 & 11 & 13 & 17\\\hline
$\xfb(p)$ & -1 & 1 & -1 & -1 & 1 & 1\\\hline
\end{tabular}
\caption{The Dirichlet character $\xfb$ evaluated on the primes in \cite{kadir:octic}.}
\label{tab:xfb}
\end{centering}
\end{table}  

It is interesting to consider the special case $g=0$, i.e. the elliptic curve \eq{y^2 = x^3 + fx,} which has $j$-invariant 1728 for all $f$. This curve is invariant under $g\to-g$, so its endomorphism ring is enlarged relative to that of a ``generic" elliptic curve; we say that a curve whose endomorphism ring is enlarged by a  twist defined over $\bbQ(\sqrt{-d})$ has complex multiplication by $\sqrt{-d}$. Thus, the $g=0$ example above has complex multiplication by $\sqrt{-4}$. Correspondingly, the modular form associated to an elliptic curve with complex multiplication is invariant under twisting by a Dirichlet character; we again say that a modular form with such a self-twist has complex multiplication. The $j$-invariants of elliptic curves over $\bbQ$ with complex multiplication have been completely classified, and there are thirteen possibilities \cite{silverman}, corresponding to the thirteen quadratic imaginary fields with class number one; this is the gateway to a rich and beautiful area of mathematics known as class field theory.

\subsection{Rational Models for Threefolds}
\label{sec:rationalThreefolds}
Before we move on, it is worth noting that most of the above discussion should also go through for CY3s, at least in principle, but of course the picture is much more complicated. Just as with elliptic curves, to discuss the modularity of a CY3 we must first select a rational model for the threefold. Thus, the modularity conjectures of \cite{RAN:fluxModularity,candelas:attractors} can more precisely be stated as conjecturing the modularity of all rational models of supersymmetric flux compactifications or rank-two attractors over $\bbQ$, and when we identify particular modular forms (as was done in \cite{RAN:fluxModularity,candelas:attractors}) we have implicitly picked a rational model. We will  shortly encounter a second family of rational models for the supersymmetric flux vacua studied in \cite{RAN:fluxModularity} that will demonstrate concretely the need for such a choice even in threefolds. 

In general, we expect the modular forms associated to different rational models of the same special point in complex structure moduli space to be related by a twist, just as with elliptic curves. However, an important difference between threefolds and elliptic curves concerns the level $N$ of the eigenform associated to a rational model of a CY3. For an elliptic curve, $N$ is simply the conductor of the curve. However, for threefolds, no precise statement is known; it is not even known whether all bad primes of the threefold must divide the level of the eigenform \cite{yui:review}. Indeed, in \cite{meyer:book} several concrete examples are given of nonrigid CY3s associated to both a weight-four eigenform and a weight-two eigenform such that the levels of the two eigenforms do not match. 

While for elliptic curves we were able to construct an infinite tower of twists, for threefolds the situation is rather more complicated. There is one lesson from elliptic curves, however, which will be quite helpful to us later on. The rescaling in Eq. \ref{eq:yRescale} is a symmetry of the underlying complex variety; however, this symmetry is broken by the choice of rational structure.  Thus, an easy way to generate different rational models for the same complex manifold is to look for symmetries of the complex moduli space that cannot be defined over $\bbQ$. 

\section{F-Theory and Elliptic Curves}
\label{sec:sen}
In \cite{RAN:fluxModularity} we conjectured that a special class of string compactifications called supersymmetric flux vacua are modular. In defining a supersymmetric flux vacuum, we start with type IIB string theory compactified to four dimensions on a CY3 $X$, and perform a special quotient of $X$ called an orientifold (see e.g. \cite{kachru:gkp}). Orientifolds of IIB string theory can be naturally reinterpreted in terms of a twelve-dimensional theory called F-theory compactified to four dimensions on an elliptically fibered fourfold $Y$ \cite{sen:orientifold}; the original threefold is a double cover of the base of the fibration. Given that the weight-two eigenforms associated to supersymmetric flux compactifications are also associated to elliptic curves, it is quite natural to hope that the elliptic fibers of the fourfold might enjoy modularity properties related to those of the threefold. 

We will see that this is the case, but first we must explain how to associate elliptic curves to supersymmetric flux compactifications, following \cite{sen:orientifold}. We begin with an elliptically fibered CY fourfold in Weierstrass form, \eq{y^2 = x^3 + f(u)x + g(u),\label{eq:weierstrasssen}} where $f$ and $g$ are functions of the coordinates $u$ of the base $\calB$ of the elliptic fibration; actually, they are sections of the bundles $-4K$ and $-6K$, respectively, where $K$ is the canonical bundle of the base, but we will not need this extra complication. We work on the locus of fourfold moduli space where $f$ and $g$ can be decomposed in terms of auxiliary functions $h(u),\eta(u)$ as \subeqs{f(u) &= C\eta(u)-3h(u)^2\label{eq:f(u)def}\\g(u) &= h(u)\left[C\eta(u)-2h(u)^2\right].} In terms of $\eta$ and $h$, the fiber over a point $u\in\calB$ has $j$-invariant \begin{subequations} \eq{j(u) = 6912\frac{\left[C\eta(u)-3h(u)^2\right]^3}{C^2\eta(u)^2\left[4C\eta(u)-9h(u)^2\right]}\label{eq:j(u)}} and discriminant \eq{\De(u) = C^2\eta(u)^2\left[4C\eta(u)-9h(u)^2\right].\label{eq:De(u)}}  \end{subequations} By analyzing monodromies around the singular locus $\De=0$, Sen argues that we have D7 branes on the locus $\eta(u)=0$, and O7 planes on the locus $h(u) = \pm\frac{2}{3}\sqrt{C\eta(u)}$. 

A particularly simple way to obtain a supersymmetric flux compactification is to take the D7 branes and O7 planes to coincide with each other; we therefore set \eq{\eta(u)=h(u)^2} so that \subeqs{f(u) &= \left(C-3\right)h(u)^2 \\ g(u) &= \left(C-2\right)h(u)^3.} Our elliptic fibration is thus \eq{y^2 = x^3 + \left(C-3\right)h(u)^2x + \left(C-2\right)h(u)^3,} with $j$-invariant \eq{j =  6912 \frac{\left(C-3\right)^3}{C^2\left(4C-9\right)}.\label{eq:j(C)}} Thus, away from the locus $h = 0$, where $\De=0$ and the fiber is singular, the $j$-invariant of the fiber is constant over the base, and depends only on the constant $C$.  

At this stage, the Weierstrass coefficients $f$ and $g$ themselves vary as we move over the base. However, this can be easily remedied. Consider the double cover of the base defined by \eq{\xi^2 - h(u) = 0.} Sen \cite{sen:orientifold} proved that this space is a CY threefold, which should be identified with the CY3 whose orientifold gives the flux compactification in the IIB picture. For our purposes, the relevant threefold is a projective variety, and hence (as long as we avoid the singular locus where $h=0$), we can projectively set $\xi=1$, which in turn sets $h=1$. Thus, away from the discriminant locus, we have a generic elliptic fiber given by \eq{y^2 = x^3 + (C-3)x + (C-2).\label{eq:ecC}} Our generic elliptic fiber has an undetermined constant $C$, and indeed the $j$-invariant of the generic fiber depends on this constant. This simply reflects the fact that we have not completely fixed the complex structure of the fourfold; to do so, we must return to the IIB description of the flux compactification. 

Recall that, in the IIB frame, to specify a flux compactification, we need to fix a family $X$ of CY3s, a point $X_*$ in the complex structure moduli space of $X$, integral fluxes $h,f\in H^3(X,\bbZ)$, and an axiodilaton $\t\in\bbH/\sltz$. These data specify a supersymmetric flux compactification if \eq{G_3 \equiv f - \t h \in H^{2,1}\left(X_*\right).\label{eq:g3}} Thus, the fluxes $f,h$ specify a natural choice of $j$-invariant, namely $j(\t)$. In fact, in F-theory, the $j$-invariant of the generic fiber should match $j(\t)$. Comparing to Eq. \ref{eq:ecC}, we can accomplish this by setting $j(\t) = j(C)$ (where $j(C)$ is given in Eq. \ref{eq:j(C)}) and solving for $C(j)$. 

Actually, this prescription is not quite well-defined yet. As a consequence of the definition of a supersymmetric flux vacuum, the fluxes $f,h$ themselves have restricted Hodge type, and in particular they satisfy \eq{f,h\in H^{2,1}\left(X_*\right) \oplus H^{1,2}\left(X_*\right).\label{eq:fh21plus12}} We then have that the $\bbZ$-span of $f$ and $h$ also has restricted Hodge type, so that there exists a two-dimensional sublattice $H_{\text{flux}}\subset H^3(X,\bbZ)$ such that \eq{\hflux \equiv \bbZ f + \bbZ h \subset H^{2,1}\left(X_*\right) \oplus H^{1,2}\left(X_*\right);} this is the same sublattice which, upon application of the Hodge conjecture, is related to the modularity of $X_*$. While it is not true that each choice of fluxes in $\hflux$ yields a physically acceptable supersymmetric flux compactification, it is true that, for any such choice, there exists some choice of $\t$ such that Eq. \ref{eq:g3} is satisfied. In general, different choices of $f,h$ will have a $\t$ related by an element of $\operatorname{GL}(2,\bbZ)$, instead of $\sltz$, and thus $j(\t)$ will depend on the choice of fluxes. To resolve this issue, we simply choose $f,h$ to be primitive lattice vectors, and select the $\t$ associated to them; all such choices are related by $\sltz$, so there is no ambiguity in this choice of $j(\t)$. 

Having chosen $j(\t)$, we can now return to Eq. \ref{eq:ecC}. Setting $j(\t)=j(C)$, where $j(C)$ is given in Eq. \ref{eq:j(C)}, and solving for $C$ yields three solutions for $C(j)$; these solutions are too long to be written down explicitly, but for rational $j$ they are in general not rational, or even real. 

We would like to use the elliptic curves in Eq. \ref{eq:ecC} to explain the modularity of supersymmetric flux compactifications. For this to work, we need the elliptic curve to be defined over $\bbQ$ whenever $X_*$ is. However, in some sense the elliptic curves are as ``bad" as could possibly be expected. Even if $j(\t)$ is rational for some $X_*$, $C(j)$ need not be, and so the elliptic curve in Eq. \ref{eq:ecC} need not be either. We will spend the remainder of this paper applying this framework to the example which we studied in \cite{RAN:fluxModularity}. We will see that the appropriate $j$-invariants are such that Eq. \ref{eq:ecC} is defined over $\bbQ$ whenever the threefold is, and therefore we have a hope of interpreting the modularity of the threefold in terms of the F-theory geometry. 

\section{A Case Study: The CY3 in $\bbP(1,1,2,2,2)$}
\label{sec:main}
To combine all of the above ideas, we will return to the example studied in \cite{RAN:fluxModularity}: the (mirror of the) octic Calabi-Yau hypersurface in $\bbP(1,1,2,2,2)$, defined as a resolution of a quotient of the projective polynomial \eq{x_1^8 + x_2^8 + x_3^4 + x_4^4 + x_5^4 - 8\psi x_1x_2x_3x_4x_5 - 2\phi x_1^4x_2^4=0.\label{eq:11222}} Along the locus $\psi=0$, these threefolds support a continuous family of supersymmetric flux vacua, parameterized by the unfixed coordinate $\phi$ \cite{dewolfe:11222}. The fourfold used to construct the F-theory geometry dual to these flux compactifications has been constructed explicitly in \cite{Collinucci:11222} as a toric variety; we will not need to use this construction, and will instead be able to work directly with the Weierstrass model in Eq. \ref{eq:weierstrasssen}.

Flux modularity implies that, subject to the usual assumptions, for each rational value of $\phi$, the underlying threefold is modular, and associated to the Tate twist of a weight-two eigenform. In \cite{RAN:fluxModularity}, we studied several such choices of $\phi$, and found evidence of modularity in each case. In this section, we will reinterpret these results in the context of the elliptic curves introduced above. Before we do so, however, we should first discuss the choice of rational model that underpinned our earlier discussion of the modularity of these varieties.

\subsection{Two Rational Models and Their Modularity}
As discussed above, whenever we discuss the modularity of a complex manifold, we must first pick a rational model for the variety. However, given the explicit form of the projective hypersurface in Eq. \ref{eq:11222}, as long as $\phi$ is rational there is an obvious choice of rational model: the projective hypersurface itself defines a perfectly good rational model. This is the rational model whose modularity we studied in \cite{RAN:fluxModularity}. 

However, there is another very natural choice of rational model for the same varieties. We saw in Section \ref{sec:rationalThreefolds} that a particularly easy way to generate different rational models is to look for symmetries of the complex manifold that are defined over fields bigger than $\bbQ$. As it turns out, the octic hypersurface has such a symmetry. By rescaling, for instance, $x_1 \to e^{i\pi/4}x_1$, we generate a symmetry on moduli space that takes $(\psi,\phi)\to(e^{i\pi/4}\psi,-\phi)$.

On the locus $\psi = 0$, where we are studying supersymmetric flux compactifications, this is simply a $\bbZ_2$ symmetry generated by $\phi\to-\phi$. However, the symmetry map is defined over the cyclotomic field $\bbQ(\zeta_8)$ so, although $\phi\to-\phi$ is a symmetry of the moduli space, it is not a symmetry of the rational varieties. Thus, the two families \subeqs{ x_1^8 + x_2^8 + x_3^4 + x_4^4 + x_5^4 - 2\phi x_1^4x_2^4=0, \label{eq:rmmp}  \\ x_1^8 + x_2^8 + x_3^4 + x_4^4 + x_5^4 + 2\phi x_1^4x_2^4=0,\label{eq:rmpp}} are isomorphic as complex varieties, but not as rational models, so that Eqs. \ref{eq:rmmp} and \ref{eq:rmpp} are two different rational models for the same point in moduli space. For simplicity, we will take $\phi\ge0$ for the remainder of the paper. 

In \cite{RAN:fluxModularity}, we studied the modularity of the rational model in Eq. \ref{eq:rmmp}. However, we could just as well have studied the modularity of Eq. \ref{eq:rmpp}. The results of doing so are listed alongside the modularity results of \cite{RAN:fluxModularity} in Table \ref{tab:phi+-comparison}.\footnote{The modular forms associated to Eq. \ref{eq:rmmp} here differ slightly from those in version 2 of \cite{RAN:fluxModularity}. For more details on this discrepancy, refer to version 3 of \cite{RAN:fluxModularity}.}$^{,}$\footnote{Here and throughout, we refer to individual eigenforms by their LMFDB labels \cite{lmfdb}. These labels indicate the level and weight of each eigenform.} The argument underlying flux modularity is phrased purely in terms of the complex structure of the underlying complex manifold, and thus we expect all rational models for a threefold over $\bbQ$ admitting a supersymmetric flux compactification to be modular (but of course different rational models will in general be related to different modular forms). This intuition is supported by the results of Table \ref{tab:phi+-comparison}, where we find eigenforms associated to both of the rational models for each value of $\phi$. 

\begin{table}[h]
\begin{centering}
\begin{tabular}{|c|c|c|c|}\hline
$\phi$ & Form from Eq. \ref{eq:rmmp} & Form from Eq. \ref{eq:rmpp} & Relative Twist \\\hline
0 & 64.2.a.a &  64.2.a.a & $\xfb$\\\hline
1/2 & 24.2.a.a & 48.2.a.a & $\xfb$ \\\hline
3/5 & 400.2.a.e & 200.2.a.c & $\xfb$ \\\hline
11/8 & 57.2.a.c & 912.2.a.b & $\xfb$ \\\hline
2 & 192.2.a.a &  192.2.a.c & $\xfb$ \\\hline
3 & 32.2.a.a & 32.2.a.a & $\xfb$\\\hline
7 & 48.2.a.a & 24.2.a.a & $\xfb$ \\\hline
9 & 40.2.a.a &  80.2.a.a & $\xfb$ \\\hline
\end{tabular}
\caption{The weight-two eigenforms associated to Eqs. \ref{eq:rmmp} and \ref{eq:rmpp} for the values of $\phi$ in \cite{RAN:fluxModularity}, as well as the relative twist between the two forms. We see that, for all of these points, the relative twist is $\xfb$.}
\label{tab:phi+-comparison}
\end{centering}
\end{table}

As different rational models, it is reasonable to expect that the modular forms associated to Eqs. \ref{eq:rmmp} and \ref{eq:rmpp} might be related by a twist. However, the coordinate redefinition that takes us between these two rational models is not defined over a quadratic imaginary field; it is therefore unclear how we should pick the twist. Table \ref{tab:phi+-comparison} provides a surprisingly simple answer. For each $\phi$ we studied, the modular forms associated to Eq. \ref{eq:rmpp} and \ref{eq:rmmp} are related by twisting by $\xfb$, exactly the same Dirichlet character we encountered in Section \ref{sec:xfb}! This striking coincidence, which we will shortly be able to explain, is one hint that the modularity of these varieties should be able to be interpreted in terms of an auxiliary geometric object that transforms more simply.

\subsection{The Elliptic Curves}
\begin{figure}[h]
\begin{center}
\includegraphics{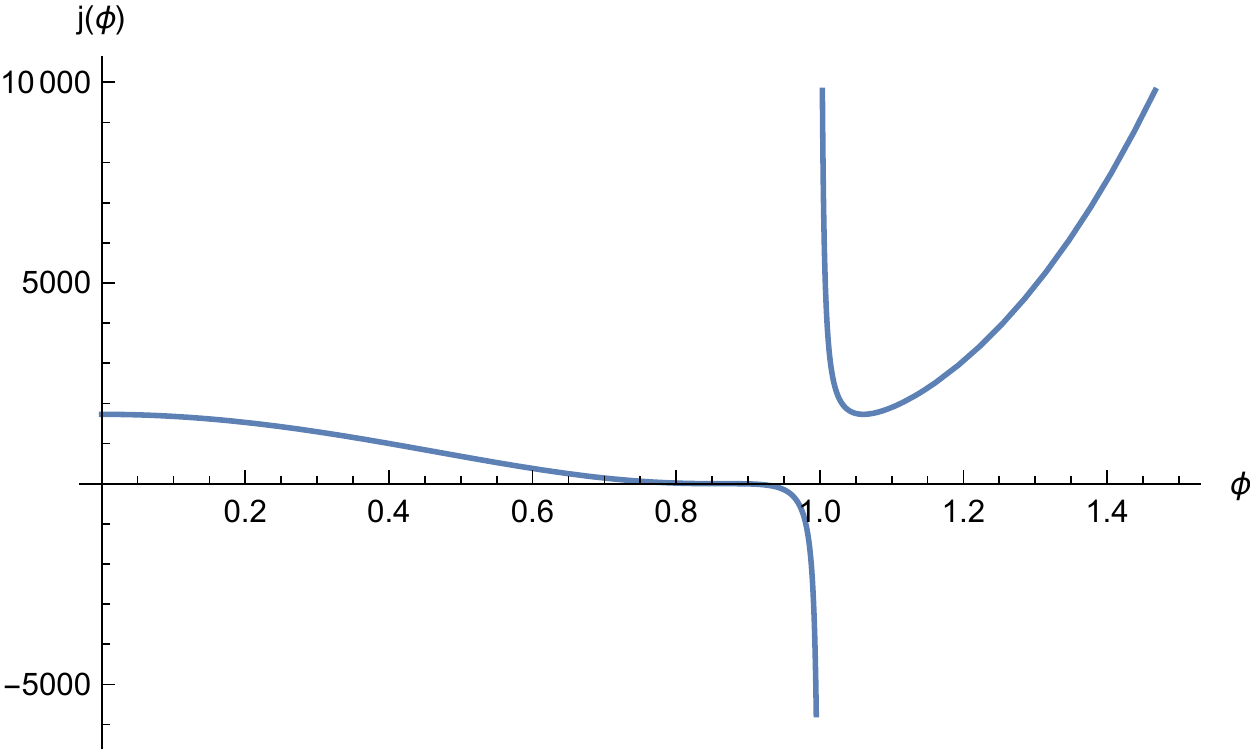}
\caption{The $j$-invariant of the elliptic curves associated to supersymmetric flux vacua, parameterized by $\phi$ as in Eq. \ref{eq:j(phi)}.}
\label{fig:j(phi)}
\end{center}
\end{figure}

Let us now apply the discussion of Section \ref{sec:sen} to this example. The first thing we need to do is compute the dilaton associated to the flux compactifications. Full details are left to Appendix \ref{sec:periods}, but we will sketch the calculation here. In terms of the fluxes $f,h$ and the periods $\Pi$, $\t$ is given as a function of $\phi$ as \cite{dewolfe:11222} \eq{\t(\phi) = \frac{f\cdot\d_\psi\Pi}{h\cdot\d_\psi\Pi}.} Using the primitive flux vectors \eq{f = \left(\begin{array}{c}2\\0\\-2\\-2\\1\\-1\end{array}\right),\ \ \ h = \left(\begin{array}{c}2\\0\\0\\0\\0\\1\end{array}\right)} and the analytic properties of the periods $\Pi$ described in \cite{candelas:2param1}, it is straightforward to compute that \eq{\t(\phi) = i\frac{u_{-1/2}(-\phi)}{u_{-1/2}(\phi)},} where $u_{-1/2}(\phi)$ is defined in \cite{candelas:2param1} and admits a series expansion given Eq. \ref{eq:u-1/2series}. Finally, comparing with the periods of the famous Legendre family of elliptic curves, we compute \eq{j(\phi) \equiv j\left[\t(\phi)\right] = 64\frac{\left(4\phi^2-3\right)^3}{\phi^2-1}. \label{eq:j(phi)}} We have graphed $j(\phi)$ in Figure \ref{fig:j(phi)}. Just from $j(\phi)$, we already see something interesting: whenever $\phi$ is rational, so is $j(\phi)$. The converse is not necessarily true: although each rational value of $j(\phi)$ is realized for some $\phi$, in general the appropriate value of $\phi$ will not be rational. 

Returning to Eq. \ref{eq:ecC}, we can set $j(C)$ equal to $j(\phi)$ to find $C(\phi)$. Doing so yields three solutions: \subeqs{C_1(\phi) &= \frac{9}{4\phi^2}\label{eq:C1} \\ C_2(\phi) &= \frac{18 \left(4 \phi^4-13 \phi^2-\sqrt{16 \phi^8-40 \phi^6+33 \phi^4-9 \phi^2}+9\right)}{64 \phi^4-144 \phi^2+81}\label{eq:C2} \\ C_3(\phi) &= \frac{18 \left(4 \phi^4-13 \phi^2+\sqrt{16 \phi^8-40 \phi^6+33 \phi^4-9 \phi^2}+9\right)}{64 \phi^4-144 \phi^2+81}.\label{eq:C3}}  Somewhat miraculously, $C_1(\phi)$ is a rational function of $\phi$! Thus, if we set $C = C_1(\phi)$ in Eq. \ref{eq:ecC}, we find a family of elliptic curves, \eq{y^2 = x^3 + \left(\frac{9}{4\phi^2}-3\right)x + \left(\frac{9}{4\phi^2}-2\right)\label{eq:ecPhiBad}.}

Before we move on, it is worth considering what we have done.  We have ``derived" a one-parameter family of elliptic curves, parameterized by our modulus $\phi$, such that the $j$-invariant matches the $j$-invariant of the flux compactifications on the mirror of the CY hypersurface in $\bbP(1,1,2,2,2)$, and moreover such that the elliptic curves are defined over $\bbQ$ whenever $\phi$ is rational. This means that for each rational value of $\phi$, we get a rational elliptic curve, and therefore a weight-two eigenform. In general, these eigenforms will be twists of the eigenforms we found in \cite{RAN:fluxModularity}. In some sense, this is the best outcome that we have any right to expect. After all, the discussion in Section \ref{sec:sen} is phrased purely in the language of complex geometry, without any reference to a rational structure in either the threefold or the fourfold. In this sense, it is surprising that the elliptic curves are rational in the first place, and not surprising that we didn't land on the ``right" rational model. After all, all rational models for the same elliptic curve are isomorphic over $\bbC$, so to F-theory as it is usually understood they should be identical!

  However, we can do even better.  As $\phi\to0$ our curve becomes badly behaved: the Weierstrass coefficients $f$ and $g$ (which after projectivizing $h$ away depend only on $\phi$) diverge. On the other hand, $j(\phi)$ is perfectly well-defined in this limit, so we don't have a genuine singularity. Instead, we have simply picked bad coordinates. The Weierstrass variables $x,y$ aren't a good description of the geometry in the $\phi\to0$ limit. 

We can fix this by simply performing a $\phi$-dependent rescaling of $y$. As discussed in Section \ref{sec:xfb}, under the transformation $y\to a^{1/2}y$, the Weierstrass form $y^2 = x^3 + fx + g$ gets transformed to $y^2 = x^2 + a^2fx + a^3g$, so our Weierstrass form becomes \eq{y^2 = x^3 + \left(\frac{9}{4\phi^2}-3\right)a^2x + \left(\frac{9}{4\phi^2}-2\right)a^3.\label{eq:ellCa}} Now we will need to cheat a little bit. We identified the point $\phi=0$ with the eigenform 64.2.a.a, which is associated to the elliptic curve \eq{y^2 = x^3 + x.\label{eq:64.2.a.aCurve}} We can pick $a$ such that, in the $\phi\to0$ limit, Eq. \ref{eq:ellCa} becomes Eq. \ref{eq:64.2.a.aCurve} if we find an $a(\phi)$ such that \begin{subequations}\eq{\lim_{\phi\to0} \left[\left(\frac{9}{4\phi^2}-3\right)a^2\right]  &= 1 \\ \lim_{\phi\to0} \left[\left(\frac{9}{4\phi^2}-2\right)a^3\right]  &= 0.}\label{eq:phito0limits}\end{subequations} The easiest way to accomplish this is to set \eq{a(\phi) = \frac{2\phi}{3}\label{eq:a=2phi/3}} so that our family of elliptic curves becomes \eq{y^2 = x^3 + \left(1-\frac{4\phi^2}{3}\right)x + \left(\frac{2\phi}{3} - \frac{16\phi^3}{27}\right).\label{eq:family}} We will call the curve associated to each $\phi$ $\e_\phi$.

It is straightforward to check, with the help of e.g. LMFDB \cite{lmfdb}, that this family of elliptic curves correctly reproduces the modular forms associated to Eq. \ref{eq:rmmp} for each value of $\phi$ studied in \cite{RAN:fluxModularity}. In Table \ref{tab:family} we have listed the LMFDB label of the elliptic curve for each of these values of $\phi$, as well as its Weierstrass coefficients and the associated modular form; we find excellent agreement with Table \ref{tab:phi+-comparison}. 

\begin{table}[h]
\begin{centering}
\begin{tabular}{|c|c|c|c|}\hline
$\phi$ & LMFDB label & Weierstrass Coefficients of $\ep$ & Associated form \\\hline
0 & 64.a4 & $\left\{1,0\right\}$ & 64.2.a.a \\\hline
$1/2$ & 24.a5 & $\left\{2/3, 7/27\right\} $ & 24.2.a.a  \\\hline
$3/5$ & 400.e4 &$\left\{13/25, 34/125\right\} $ & 400.2.a.e  \\\hline 
11/8 & 57.c3 & $ \left\{-73/48, -539/864 \right\} $ & 57.2.a.c   \\\hline
$2$ & 192.a3& $\left\{ -13/3, -92/27\right\}$ & 192.2.a.a  \\\hline
3 & 32.a1  & $\left\{-11, -14\right\} $ & 32.2.a.a  \\\hline
7 & 48.a2 & $\left\{-193/3, -5362/27\right\}$ & 48.2.a.a  \\\hline
$9$ &40.a1 & $\left\{ -107, -426\right\}$ & 40.2.a.a  \\\hline
\end{tabular}
\caption{Elliptic curves from Eq. \ref{eq:family} and their associated modular forms for the $\phi$ points listed in Table \ref{tab:phi+-comparison}. The notation $\left\{f,g\right\}$ means the elliptic curve $y^2 = x^3 + fx + g$.}
\label{tab:family}
\end{centering}
\end{table}

What about our other rational model, Eq. \ref{eq:rmpp}? By picking $a\sim\phi$ in Eq. \ref{eq:a=2phi/3}, we have explicitly broken the $\phi\to-\phi$ symmetry of our complex structure moduli space. We could just have easily have picked \eq{a = -\frac{2}{3}\phi\label{eq:a=-2phi/3}} to find the family \eq{y^2 = x^3 + \left(1-\frac{4\phi^2}{3}\right)x - \left(\frac{2\phi}{3} - \frac{16\phi^3}{27}\right),\label{eq:family-}} which is of course simply the image of Eq. \ref{eq:family} under $\phi\to-\phi$. We will call these elliptic curves $\epp$. Again, we can check whether the forms associated to these curves match those associated to Eq. \ref{eq:rmpp} in Table \ref{tab:phi+-comparison}; the results are summarized in Table \ref{tab:family-}, and again we find matching for all $\phi$. 
\begin{table}[h]
\begin{centering}
\begin{tabular}{|c|c|c|c|}\hline
$\phi$ & LMFDB label & Weierstrass Coefficients of $\epp$ & Associated form \\\hline
0 & 64.a4 & $\left\{1,0\right\}$ & 64.2.a.a \\\hline
$1/2$ & 48.a5 & $\left\{2/3, -7/27\right\} $ & 48.2.a.a  \\\hline
$3/5$ & 200.c4 &$\left\{13/25, -34/125\right\} $ & 200.2.a.c  \\\hline 
11/8 & 912.b3 & $ \left\{-73/48, 539/864 \right\} $ & 912.2.a.b   \\\hline
$2$ & 192.c3 & $\left\{ -13/3, 92/27\right\}$ & 192.2.a.c  \\\hline
3 & 32.a2  & $\left\{-11, 14\right\} $ & 32.2.a.a  \\\hline
7 & 24.a2 & $\left\{-193/3, 5362/27\right\}$ & 24.2.a.a  \\\hline
$9$ & 80.a1 & $\left\{ -107, 426\right\}$ & 80.2.a.a  \\\hline
\end{tabular}
\caption{Elliptic curves from Eq. \ref{eq:family-} and their associated modular forms for the $\phi$ points listed in Table \ref{tab:phi+-comparison}. The notation $\left\{f,g\right\}$ means the elliptic curve $y^2 = x^3 + fx + g$.}
\label{tab:family-}
\end{centering}
\end{table}

Thus, we have two rational models for the threefolds in Eqs. \ref{eq:rmmp} and \ref{eq:rmpp}, and two rational families of elliptic curves in Eqs. \ref{eq:family} and \ref{eq:family-}. For the $\phi$ points studied in \cite{RAN:fluxModularity}, we have checked that the modular forms associated to the curves in \ref{eq:family} exactly match those associated to Eq. \ref{eq:rmmp}, and similarly the forms associated to Eq. \ref{eq:family-} match those associated to Eq. \ref{eq:rmpp}. In fact, we have checked many more points. For all $\phi = \frac{n}{d}$ where $1\le n,d< 2,000$ (in lowest terms), we have checked that these two sets of families continue to match each other: the Fourier coefficients of the modular form associated to the elliptic curve are compatible with the $\zeta$-function factorizations in \cite{kadir:octic} and version 3 of \cite{RAN:fluxModularity}. This constitutes 2,431,574 examples for each set of rational models. Thus, we conjecture that the modular forms associated to Eqs. \ref{eq:rmmp} and \ref{eq:rmpp} are given by the forms associated to the elliptic curves associated to the modular forms in \ref{eq:family} and \ref{eq:family-}, respectively, for all rational $\phi$. We therefore see that in some sense the arithmetic structure associated to these flux compactifications varies analytically with $\phi$. 

This is our main result. By ``cheating" at one value of $\phi$ to fix the complex structure of the F-theory fourfold, we have managed to reproduce the modularity of the CY hypersurface in $\bbP(1,1,2,2,2)$. Thus, not only does the F-theory prescription of \cite{sen:orientifold} describe the complex geometry of the IIB threefold, it also describes its arithmetic geometry. As emphasized before, there was no real reason for this to happen: the elliptic fibers of the fourfold are a priori not defined over $\bbQ$ at all, but instead over a function field. For the supersymmetric flux compactifications considered here, these complications are all avoided, and we actually find a simple family of rational elliptic curves.

Of course, all of this is subject to the somewhat artificial rescaling in Eqs. \ref{eq:a=2phi/3} and \ref{eq:a=-2phi/3}. The need for such a rescaling is a crucial step in the argument, and is why the relationship between the modularity of the flux compactification and the F-theory geometry is more subtle than was anticipated in \cite{candelas:attractors}.  These rescalings are simple quadratic twists of the generic fiber, so they twist the modular form associated to each $\phi$ by a ($\phi$-dependent) Dirichlet character. We picked this twist by matching on to the modular form we wanted at $\phi=0$, so we only got the ``right" rational models out after putting one model in by hand. However, this was simply a choice; we could have picked a different twist, or even no twist at all. Different choices of $a(\phi)$ would presumably yield the modular forms associated to different rational models for the underlying threefold. There is quite possibly a rational model for the CY3 in $\bbP(1,1,2,2,2)$ whose associated eigenforms match those of Eq. \ref{eq:ecPhiBad}, but it is not the rational model we considered in \cite{RAN:fluxModularity}. It would be quite enlightening to see if one could engineer the rational models for the threefold from the rational model for the elliptic curves, but we will not attempt to do so here.

\subsection{Bigger Number Fields}
So far we have restricted ourselves to rational values of $\phi$, to compare with \cite{kadir:octic,RAN:fluxModularity}. However, with the conjecture of the previous section, we can now \textit{predict} the automorphic forms associated to the CY3 in $\bbP(1,1,2,2,2)$ for certain irrational values of $\phi$. Indeed, whenever $\phi$ is in a number field $K$, Eq. \ref{eq:family} provides an elliptic curve defined over $K$ (and with  $j$-invariant also in $K$). For $K$ such that the modularity of elliptic curves over $K$ is well-understood, we can therefore associate an automorphic form to the elliptic curve $\ep$ exactly as we did before, and by the same logic as above we can conjecture that the rational model Eq. \ref{eq:rmmp} should be associated to the same automorphic form. For simplicitly, in this section we will focus on Eq. \ref{eq:family}, and correspondingly the threefolds in Eq. \ref{eq:rmmp}, but the exact same reasoning applies equally well to Eqs. \ref{eq:family-} and \ref{eq:rmpp}.

It is known that elliptic curves over real quadratic fields are associated to Hilbert modular forms. Thus, when $\phi$ is in such a field, we can find an elliptic curve from Eq. \ref{eq:family}, and plug it in to LMFDB \cite{lmfdb} to find one or more Hilbert eigenforms. Some examples are shown in Table \ref{tab:realQuadratic}. While we have not computed the $\zeta$-functions of these threefolds over such number fields, a framework for doing so was laid out in e.g. \cite{straten:hilbert,schutt:hilbert}; we predict that applying that framework to this example will reproduce the Hilbert modular forms listed in the table.

\begin{table}[h]
\begin{centering}
\begin{tabular}{|c|c|c|c|}\hline
$\phi$ & Weierstrass coefficients of $\ep$  & LMFDB Label & Associated Hibert Form(s) \\\hline
$\sqrt{2}$ & $\left\{-\frac{5}{3},-\frac{14}{27}\sqrt{2}\right\}$ &  2.2.8.1-256.1-c4 & 2.2.8.1-256.1-c \\\hline
$\sqrt{2}+1$ & $\left\{-3-\frac{8}{3}\sqrt{2},-\frac{94}{27}-\frac{62}{27}\sqrt{2}\right\}$ &  2.2.8.1-1024.1-l3 & 2.2.8.1-256.1-l, 2.2.8.1-256.1-p \\\hline
$\frac{1}{2}\sqrt{2}$ & $\left\{\frac{1}{3},\frac{5}{27}\sqrt{2}\right\}$ &  2.2.8.1-1024.1-g1 & 2.2.8.1-256.1-g, 2.2.8.1-256.1-i \\\hline
$\sqrt{3}$ & $\{-3,-\frac{10}{9}\sqrt{3}\}$ & 2.2.12.1-1024.1-r3 & 2.2.12.1-1024.1-d, 2.2.12.1-1024.1-r \\\hline
$\frac{1}{2}\sqrt{3}$ & $\{0, \frac{1}{9}\sqrt{3}\}$ & 2.2.12.1-256.1-c1 & 2.2.12.1-256.1-c \\\hline
$-\frac{1}{2}\sqrt{3}$ & $\{0, -\frac{1}{9}\sqrt{3}\}$ & 2.2.12.1-256.1-c2 & 2.2.12.1-256.1-c \\\hline
$\frac{1}{4}\sqrt{15}$ &  $\{-\frac{1}{4},\frac{1}{36}\sqrt{15}\}$ & 2.2.60.1-8.1-b1 & 2.2.60.1-8.1-b, 2.2.60.1-8.1-d \\\hline
\end{tabular}
\caption{The elliptic curves obtained from Eq. \ref{eq:family} for $\phi$ valued in real quadratic fields. For each $\phi$, we give the Weierstrass coefficients of the elliptic curve, its LMFDB label, and its associated Hilbert eigenform(s). As before, the notation $\left\{f,g\right\}$ corresponds to the Weierstrass form $y^2 = x^3 + fx + g$.}
\label{tab:realQuadratic}
\end{centering}
\end{table}

Similarly, elliptic curves over imaginary quadratic fields are associated to Bianchi modular forms. In Table \ref{tab:quadImag}, we have listed several examples of quadratic imaginary $\phi$, as well as their associated Bianchi modular forms. To the best of our knowledge, there are no known examples of modular threefolds over quadratic imaginary fields. In light of the above discussion, these values of $\phi$ provide natural candidates for the first such examples. 

Finally, we note that elliptic curves over all CM number fields (such as e.g. the cyclotomic fields) were proven to be potentially modular in \cite{taylor:CMfields}. As the modularity of such elliptic curves becomes better understood, Eq. \ref{eq:family} provides a natural way to associate (Tate twists of) whatever automorphic objects get associated to these elliptic curves to threefolds. 

\begin{table}[h]
\begin{centering}
\begin{tabular}{|c|c|c|c|}\hline
$\phi$ & Weierstrass coefficients of $\ep$  & LMFDB Label & Associated Bianchi Form(s) \\\hline
$i$ & $\{\frac{7}{3},\frac{34}{27}i\}$ & 2.0.4.1-1024.1-a4 & 2.0.4.1-1024.1-a \\\hline
$i+1$ & $\{-\frac{8}{3}i+1,-\frac{14}{27}i+\frac{50}{27}\}$ & 2.0.4.1-160.1-a2 & 2.0.4.1-160.1-a, 2.0.4.1-160.2-a \\\hline
$2i$ & $\{\frac{19}{3},\frac{164}{27}i\}$ & 2.0.4.1-6400.2-f4 & 2.0.4.1-6400.2-f \\\hline
$\frac{1}{2}i$ & $\{\frac{4}{3},\frac{11}{27}i\}$ & 2.0.4.1-6400.2-e7 & 2.0.4.1-6400.2-e \\\hline
$2i+1$ & $\{-\frac{16}{3}i+5,\frac{68}{27}i+\frac{194}{27}\}$ & 2.0.4.1-512.1-a1 & 2.0.4.1-512.1-a, 2.0.4.1-512.1-b \\\hline
$i\sqrt{2}$ & $\{\frac{11}{3},\frac{50}{27}i\sqrt{2}\}$ & 2.0.8.1-2304.2-d5 & 2.0.8.1-2304.2-d, 2.0.8.1-2304.2-f \\\hline
$i\sqrt{2}+1$ & $\{-\frac{8}{3}i\sqrt{2} + \frac{7}{3}, \frac{2}{27}i\sqrt{2}+\frac{98}{27}\}$ & 2.0.8.1-3072.2-b1 & 2.0.8.1-3072.1-b, 2.0.8.1-3072.2-b \\\hline
\end{tabular}
\caption{The elliptic curves obtained from Eq. \ref{eq:family} for $\phi$ valued in imaginary quadratic fields. For each $\phi$, we give the Weierstrass coefficients of the elliptic curve, its LMFDB label, and its associated Bianchi eigenform(s). As before, the notation $\left\{f,g\right\}$ corresponds to the Weierstrass form $y^2 = x^3 + fx + g$.}
\label{tab:quadImag}
\end{centering}
\end{table}

\section{Conclusion}
\label{sec:conclusion}
We have argued that the F-theory construction of \cite{sen:orientifold}, when applied to the one-parameter family of supersymmetric flux compactifications on the CY hypersurface in $\bbP(1,1,2,2,2)$, yields a one-parameter family of elliptic curves that reproduces the modular forms associated to the underlying Calabi-Yau threefold. In this way, we find a direct relationship between the physical flux compactification and the arithmetic properties of the underlying threefold. 

Nevertheless, we have not \textit{proven} that the threefold is modular, and associated to the same eigenform as the generic fiber.  Doing this would involve computing the $\zeta$-function of the threefold directly from the Weierstrass form of the fourfold. The CY3 in $\bbP(1,1,2,2,6)$, which also admits a one-parameter family of supersymmetric flux compactifications \cite{dewolfe:11222,kachru:fluxthreefolds}, provides a particularly natural candidate for this calculation.  One advantage it enjoys over the present model is the relative simplicity of its fourfold lift in F-theory.

Even if the calculation of the $\zeta$-function could be done, we still would not have a satisfactory physical explanation of the modularity of flux vacua. To compute the Fourier coefficients of a modular form from a projective variety, we must reduce the defining polynomial modulo various primes. Can we ever motivate this reduction in string theory? Although various references have made progress towards computing the $L$-functions of varieties from string theory (see e.g. \cite{Kadir:2010dh,schimmrigk:flux,Kondo:WorldsheetLFunctions,Kondo:WorldsheetLFunctions2}), to date no clear, string theoretic motivation for the mod-$p$ reductions of a threefold or other variety have ever been given. 

Relatedly, as emphasized throughout the paper, in physics we typically do not imagine choosing a rational model for our CY3s. What do different choices of rational model represent physically? Is there some subtle (and hitherto unidentified) physical observable which distinguishes between them? If so, then we are in some sense always picking a rational model, albeit unknowingly. The $L$-function calculation of \cite{Kondo:WorldsheetLFunctions,Kondo:WorldsheetLFunctions2} supports this idea, at least for elliptic curves, but it is still unclear how exactly to reconcile our usual intuition with these results.

Actually, even asking these questions presupposes some new conceptual advances. As demonstrated clearly by our main example, supersymmetric flux vacua need not be defined over number fields. However, to discuss modularity, we need to restrict ourselves to threefolds defined over number fields. While the Hodge conjecture implies that rank-two attractors (and therefore supersymmetric flux compactifications on threefolds with $h^{2,1}=1$) are always defined over number fields, clearly this argument does not apply to more general flux vacua. Does some still-to-be-understood physics distinguish the subset of flux vacua defined over number fields from the more generic ones? Without an affirmative answer to this question, it is hard to see how flux modularity can ever have a satisfying physical explanation. We hope that such an answer exists, and that in finding this answer we can expand the growing relationship between string theory and arithmetic geometry. 

\section*{Acknowledgements}
We are grateful to S. Kadir and W. Taylor for extremely helpful discussions, and M. Zimet for comments on an earlier version of this paper. This research of SK was supported by the NSF under grant PHY-2014215 and by a Simons Investigator Award. RN thanks C. Hewett for helpful discussions, and is funded by NSF Fellowship DGE-1656518 and an EDGE grant from Stanford University. WY is supported by the Stanford Institute for Theoretical Physics.

\appendices

\section{Deriving Eq. 4.6}
\label{sec:periods}
In this appendix we will derive Eq. \ref{eq:j(phi)}. Our starting points are the periods of the mirror of the CY3 in $\bbP(1,1,2,2,2)$, as studied in \cite{candelas:2param1}. The CY3 is equipped with a nowhere vanishing holomorphic threeform $\Om$, and the periods are defined by the integrals of the $\Om$ over some basis for $H^{3}(X)$. The holomorphic threeform $\Om$, and therefore its periods, are determined by the Picard-Fuchs (PF) equations, a complicated system of coupled nonlinear PDEs which in our example are given by \begin{equation} \label{eq:octicPFequations}
\begin{aligned}
\frac{\partial^3\Omega_{\psi,\phi}}{\partial \psi^3}-32 \psi^3 \frac{\partial^3 \Omega_{\psi,\phi}}{\partial \psi^2 \partial \phi}-96\psi^2 \frac{\partial^2 \Omega_{\psi,\phi}}{\partial \psi \partial \phi}-32 \psi \frac{\partial \Omega_{\psi,\phi}}{\partial \phi}=&0, \\
16(\phi^2-1) \frac{\partial^2 \Omega_{\psi,\phi}}{\partial \phi^2}+\psi^2 \frac{\partial^2 \Omega_{\psi,\phi}}{\partial \psi^2}+8\psi \phi \frac{\partial^2 \Omega_{\psi,\phi}}{\partial \psi \partial \phi}+3\psi \frac{\partial \Omega_{\psi,\phi}}{\partial \psi}+24 \phi \frac{\partial \Omega_{\psi,\phi}}{\partial \phi}+\Omega_{\psi,\phi}&=0,\\
(8 \psi^4+\phi+1)(8 \psi^4+\phi-1)\frac{\partial^4 \Omega_{\psi,\phi}}{\partial \psi^3 \partial \phi}+128 \psi^3(8 \psi^4+\phi) \frac{\partial^3 \Omega_{\psi,\phi}}{\partial \psi^2 \partial \phi}+50\psi^3 \frac{\partial^2 \Omega_{\psi,\phi}}{\partial \psi^2}+& \\
16 \psi^2(148 \psi^4+13 \phi) \frac{\partial \Omega_{\psi,\phi}}{\partial \psi \partial \phi}+30 \psi^2 \frac{\partial \Omega_{\psi,\phi}}{\partial \psi}+16 \psi(44 \psi^4+3 \phi) \frac{\partial \Omega_{\psi,\phi}}{\partial \phi}+2 \psi \Omega_{\psi,\phi}&=0.
\end{aligned}
\end{equation}
This system has a six-dimensional space of solutions, which matches the dimension of $H^3(X)$. 

An explicit set of solutions was found in \cite{candelas:2param1} in terms of auxiliary functions $u_l(\phi)$, defined as the solutions of the second-order ODEs \begin{equation} \label{eq:ODEul}
(\phi^2-1) \frac{d^2 u_l}{d \phi^2}-(2l-1) \phi \frac{d u_l}{d \phi}+l^2 u_l=0,
\end{equation}
for $l\in \mathbb{Z}/4$, subject to the boundary condition 
\begin{equation}
u_l(0)=\frac{2^l \pi^{\frac{1}{2}} \exp \frac{l \,\pi i}{2}}{\Gamma(1+\frac{l}{2})\Gamma(\frac{1}{2}-\frac{l}{2})}.
\end{equation}
For $l\not\in\bbZ$, which is the case that we will be the most interested in, $u_l(\phi)$ and $u_l(\phi)$ are linearly independent, and span the solution space; for $\phi>1$, they are given explicitly as \cite{candelas:2param1} \begin{equation}
\begin{aligned}
u_l(\phi)&=(2 \phi)^l\,_2F_1(-\frac{l}{2},\frac{1-l}{2};1;\frac{1}{\phi^2}),\\
u_l(-\phi)&=\frac{\sin l \pi}{\pi}(2 \phi)^l \sum_{n \geq 0}^\infty \frac{(-\frac{l}{2})_n(\frac{1-l}{2})_n}{(n!)^2} \frac{1}{\phi^{2n}}\cdot \Big[-\log(-\phi^2)+\psi^{(0)}(-\frac{l}{2}+n)\\
&+\psi^{(0)}(\frac{1-l}{2}+n)-2\psi^{(0)}(n+1) \Big]-\cos l\pi (2 \phi)^l\,_2F_1(-\frac{l}{2},\frac{1-l}{2};1;\frac{1}{\phi^2}).
\end{aligned}
\end{equation}

In the domain \begin{equation}
\Big| \frac{8 \psi^4}{\phi \pm 1} \Big| <1,
\end{equation}
\cite{candelas:2param1} provides an expansion for a basis $\varpi_j(\psi,\phi)$ (for $j\in0,1,\cdots,5$) of periods in terms of the $u_l$: \begin{equation} \label{eq:expansionvarpij}
\varpi_j(\psi,\phi)=-\frac{1}{4} \sum_{m=1}^\infty \frac{(-1)^m (-1)^{mj/4} \Gamma(\frac{m}{4}) }{\Gamma(m) \Gamma^3(1-\frac{m}{4})} (2^{12} \psi^4)^{\frac{m}{4}} u_{-\frac{m}{4}}((-1)^j \phi).
\end{equation}

The fluxes are elements of the integral cohomology $H^3(X,\bbZ)$, so to study them we must express the periods in an integral symplectic basis. An integral symplectic basis for $H_3(X,\bbZ)$ is a basis $\{A_a, B_a\}$ of cycles (for $a\in0,\cdots,h^{2,1}$) such that \eq{A_a\cdot A_b = 0 = B_a \cdot B_b, \ \ \ A_a \cdot B_b = \de_{ab}.}  We write the periods as \eq{\mathcal{G}_a = \int_{A_a}\Om, \ \ \ z_a = \int_{B_a}\Om.} It is conventional to package these periods into a column vector as \eq{\Pi = \left(\mathcal{G}_0, \mathcal{G}_1, \mathcal{G}_2, z_0, z_1, z_1\right)^\top.} We also write the Poincare dual integral symplectic basis of threeforms as \eq{\b = \left(\b_0,\b_1,\b_2,\a_0,\a_1,\a_2\right),} where $\a$ is dual to $A$ and $\b$ is dual to $B$. Up to an irrelevant overall constant, we have \begin{subequations} \eq{\Pi = S \varpi,} where \cite{candelas:2param1} \begin{equation}
S= (2 \pi i)^3 \cdot
\left(
\begin{array}{cccccc}
 -1\,\,\,\,\, & 1 & 0 & 0 & 0 & 0 \\
 1 & 0 & 1 & -1\,\,\,\,\,  & 0 & -1 \,\,\,\,\,  \\
 \frac{3}{2} & 0 & 0 & 0 & -\frac{1}{2} \,\,\,\,\,  & 0 \\
 1 & 0 & 0 & 0 & 0 & 0 \\
 -\frac{1}{4}\,\,\,\,\,  & 0 & \frac{1}{2} & 0 & \frac{1}{4} & 0 \\
 \frac{1}{4} & \frac{3}{4} & -\frac{1}{2} \,\,\,\,\,  & \frac{1}{2} & -\frac{1}{4} \,\,\,\,\,  & \frac{1}{4} \\
\end{array}
\right).
\end{equation}
\end{subequations}
We can then write the holomorphic threeform itself as \eq{\Om = \b S\varpi.} 

Let us now turn to the $\psi=0$ locus, where we expect to find supersymmetric flux vacua \cite{dewolfe:11222}. On this locus, the periods (as given in Eq. \ref{eq:expansionvarpij}) vanish. We get around this by defining a rescaled threeform \eq{\Om^R = \frac{1}{\psi} \Om,} with rescaled periods $\varpi_j^R$ and $\Pi^R$. We have \subeqs{\varpi_j^R(\phi) &= -\frac{1}{4}\left\{-\frac{\left(-1\right)^{j/4}\Gamma\left(\frac{1}{4}\right)}{\Gamma\left(\frac{3}{4}\right)^3} u_{-1/4}\left[\left(-1\right)^j\phi\right]+ 64\frac{\left(-1\right)^{j/4}}{\Gamma\left(\frac{1}{2}\right)^2}\psi u_{-1/2}\left[\left(-1\right)^j\phi\right] + \cdots\right\} \label{eq:varpiRexpand} \\ &= 2\frac{\left(-1\right)^{j/4}\Gamma\left(\frac{1}{4}\right)}{\Gamma\left(\frac{3}{4}\right)^3}u_{-1/4}\left[\left(-1\right)^j\phi\right],} and as before \subeqs{\Pi^R &= s\varpi^R \\ \Om^R &= \b S \varpi^R.}

We can now look for fluxes $f,h$ satisfying Eq. \ref{eq:fh21plus12}, i.e. elements of $H^3(X,\bbZ)$ that are orthogonal to $\Om^R$. Direct calculation shows that a basis for such vectors is \eq{V_1=(2, 0, 0, 0, 0, 1)^\top~\text{and}~V_2=(2,0,-2,-2,1,-1)^\top.} These are the generators of the flux lattice $H_{\text{flux}}$ defined above. From \cite{dewolfe:11222}, we have that the axiodilaton is given by \eq{\t_{\text{flux}}(\phi) = \frac{f\cdot \d_\psi\Pi^R}{h\cdot\d_\psi\Pi^R}.} Setting $f = V_2, h = V_1$ and using Eq. \ref{eq:varpiRexpand} to take the derivatives, we find that \eq{\t_{\text{flux}}(\phi) = i\frac{u_{-1/2}(-\phi)}{u_{-1/2}(\phi)},\label{eq:appendixTau}} where  \begin{subequations}
\seq{u_{-\frac{1}{2}}(\phi)&=(2 \phi)^{-\frac{1}{2}}\,_2F_1\left(\frac{1}{4},\frac{3}{4};1;\frac{1}{\phi^2}\right)\numberthis\\
u_{-\frac{1}{2}}(-\phi)&=-\frac{1}{\pi}(2 \phi)^{-\frac{1}{2}} \sum_{n \geq 0}^\infty \frac{(\frac{1}{4})_n(\frac{3}{4})_n}{(n!)^2} \frac{1}{\phi^{2n}}\cdot \Bigg[-\log\left(-\phi^2\right)+\psi^{(0)}\left(\frac{1}{4}+n\right)\\
&~~~~~~~~~~~~~~~~~~~~~~~~~~~~~~~~~~~~~~~~~~~~~ +\psi^{(0)}\left(\frac{3}{4}+n\right)-2\psi^{(0)}(n+1) \Bigg].\numberthis}
\label{eq:u-1/2series}
\end{subequations}

We now need to compute $j\left[\tau_{\text{flux}}(\phi)\right].$ However, this is quite nontrivial, given that $u_{-1/2}(-\phi)$ does not admit a simple expression. The easiest way we have found to do this calculation is to compare to the periods of the Legendre family of elliptic curves, defined by the (long) Weierstrass form \eq{y^2 = x(x-1)(x-\l).} As a function of $\lambda$, these curves have $j$-invariant \eq{j(\l) =  256 \frac{\left[\l\left(\l-1\right)-1\right]^3}{\l^2\left(\l-1\right)^2}.\label{eq:legendrej}} We will find a function $\l(\phi)$ such that the modular parameter $\t(\l)$ of these elliptic curves exactly reproduces Eq. \ref{eq:appendixTau}, which will let us find $j(\phi)$ in closed form. 

We begin with the PF equation for the Legendre family, given by \cite{carlson2017period} \begin{equation} \label{eq:picardfuchsequationlegendre}
\lambda(\lambda-1)\,\frac{d^2 \Omega}{d\lambda^2} +(2\lambda-1) \,\frac{d \,\Omega}{d \lambda}+\frac{1}{4} \,\Omega =0.
\end{equation}
In a neighborhood of $\l=0$, the periods can be written as \begin{equation}
\begin{aligned}
\pi_0(\lambda)&= \sum_{n=0}^\infty \binom{-1/2}{n}^2 \lambda^n=\, _2 F_1(\frac{1}{2},\frac{1}{2};1;\lambda),\\
\pi_1(\lambda)&=\frac{1}{\pi i} (\pi_0(\lambda) \log \lambda+h(\lambda))-\frac{\log 16}{\pi i} \pi_0(\lambda),
\end{aligned}
\end{equation}
where \cite{carlson2017period,wenzhe:lambda} \begin{equation}
h(\lambda) =\frac{1 }{2}\,\lambda+\frac{21 }{64}\,\lambda^2+\frac{185 }{768}\,\lambda^3+ \cdots.
\end{equation}
The modular parameter of these curves is given by \eq{\t_{\text{Legendre}}(\l) = \frac{\pi_1(\l)}{\pi_0(\l)}.} To match Eq. \ref{eq:legendrej} we must have that \eq{j[\t_{\text{Legendre}}(\l)] = 256 \frac{\left[\l\left(\l-1\right)-1\right]^3}{\l^2\left(\l-1\right)^2}.}

Both $\pi_0(\l)$ and $u_{-1/2}(\phi)$ are given in terms of $_2F_1$ hypergeometric functions. Moreover, there is a standard relationship between hypergeometric functions of the form \cite{HG1,HG2,HG3} \begin{equation}
_2F_1(\frac{1}{2},\frac{1}{2};1;\lambda)=(1-\lambda)^{-\frac{1}{2}}     \, _2F_1(\frac{1}{4},\frac{3}{4};1;\frac{4\lambda}{(1+\lambda)^2}).
\end{equation}
The left-hand side of this equation appears in $\pi_0(\l)$ and, subject to the identification \eq{\frac{1}{\phi^2} = \frac{4\l}{\left(\l+1\right)^2},\label{eq:l(phi)}} the hypergeometric function on the right-hand side is exactly $u_{-1/2}(\phi)$, so this equation provides a map between $\pi_0(\l)$ and $u_{-1/2}(\phi)$. Indeed, we have that \subeqs{u_{-\frac{1}{2}}(\phi) &=\lambda^{\frac{1}{4}}(1-\lambda)^{\frac{1}{2}}(1+\lambda)^{-\frac{1}{2}}\,_2F_1(\frac{1}{2},\frac{1}{2};1;\lambda)=\lambda^{\frac{1}{4}}(1-\lambda)^{\frac{1}{2}}(1+\lambda)^{-\frac{1}{2}} \pi_0(\lambda) \\ 
u_{-\frac{1}{2}}(-\phi)& =-i\,\lambda^{\frac{1}{4}}(1-\lambda)^{\frac{1}{2}}(1+\lambda)^{-\frac{1}{2}} (\pi_1(\lambda)-\pi_0(\lambda)).}

Plugging into Eq. \ref{eq:appendixTau}, we have that \eq{\t_{\text{flux}}(\phi) = \frac{\lambda^{\frac{1}{4}}(1-\lambda)^{\frac{1}{2}}(1+\lambda)^{-\frac{1}{2}} (\pi_1(\lambda)-\pi_0(\lambda))}{\lambda^{\frac{1}{4}}(1-\lambda)^{\frac{1}{2}}(1+\lambda)^{-\frac{1}{2}} \pi_0(\lambda)} = \frac{\pi_1(\l)}{\pi_0(\l)} - 1 = \t_{\text{Legendre}}(\l) - 1,} where $\l$ is determined in terms of $\phi$ by Eq. \ref{eq:l(phi)}. We then have that \eq{j\left[\t_{\text{flux}}(\phi)\right] = j\left[\t_{\text{Legendre}}(\l)\right].} Plugging Eq. \ref{eq:l(phi)} into Eq. \ref{eq:legendrej}, we find that \eq{j(\phi) = 64\frac{\left(4\phi^2-3\right)^3}{\phi^2-1},} which completes our derivation of Eq. \ref{eq:j(phi)}.

\bibliographystyle{jhep}
\bibliography{flux}

\end{document}